\documentclass[a4paper,12pt,reqno]{amsart}

\usepackage[centertags]{amsmath}
\usepackage{amsfonts}
\usepackage{euscript}
\usepackage{amssymb}
\usepackage{amsthm}
\usepackage{newlfont}
\usepackage{stmaryrd}
\usepackage{mathrsfs}
\usepackage{euscript}

\usepackage{color}

\theoremstyle{plain}
  \newtheorem{theorem}{Theorem}[section]
  
  \newtheorem{proposition}[theorem]{Proposition}
  \newtheorem{lemma}[theorem]{Lemma}
\theoremstyle{definition}

\theoremstyle{remark}
  \newtheorem{remark}[theorem]{Remark}

\numberwithin{equation}{section}

\let\al=\alpha \let\be=\beta \let\de=\delta \let\ep=\epsilon
  \let\ga=\gamma 
\let\ka=\kappa \let\la=\lambda \let\om=\omega 
\let\si=\sigma  \let\th=\theta

\let\De=\Delta   \let\Om=\Omega
\let\Th=\Theta

\newcommand{\caA}{{\mathcal A}}

\newcommand{\caD}{{\mathcal D}}
\newcommand{\caE}{{\mathcal E}}

\newcommand{\caL}{{\mathcal L}}
\newcommand{\caM}{{\mathcal M}}

\newcommand{\caO}{{\mathcal O}}

\newcommand{\scD}{{\mathscr D}}

\newcommand{\bbR}{{\mathbb R}}

\newcommand{\opunit}{\text{1}\kern-0.22em\text{l}}
\newcommand{\funit}{\mathbf{1}}

\newcommand{\bsE}{{\boldsymbol E}}

\newcommand{\bsP}{{\boldsymbol P}}

\DeclareMathAlphabet{\mathpzc}{OT1}{pzc}{m}{it}

\newcommand{\pzO}{\mathpzc{O}}

\newcommand{\rel}{\,|\,}

\newcommand{\0}{^{(0)}}
\newcommand{\1}{^{(1)}}
\newcommand{\2}{^{(2)}}
\newcommand{\tot}{_{\text{TOT}}}
\newcommand{\out}{_{\text{OUT}}}

\newcommand{\id}{\textrm{d}}

\begin{document}

\begin{center}
\noindent{\large \bf Minimum entropy production principle from
\\
 a dynamical fluctuation law} \\

\vspace{15pt}

{\bf Christian Maes\footnote{email: {\tt Christian.Maes@fys.kuleuven.be}}}\\
Instituut voor Theoretische Fysica, K.U.Leuven\\
\vspace{5pt} and\\ \vspace{5pt} {\bf Karel
Neto\v{c}n\'{y}\footnote{email: {\tt netocny@fzu.cz}}}\\
Institute of Physics AS CR, Prague
\end{center}

\vspace{20pt} \footnotesize \noindent {\bf Abstract: } The minimum
entropy production principle provides an approximative variational characterization of
close-to-equilibrium stationary states, both for macroscopic systems and for stochastic models.
Analyzing the fluctuations of the empirical distribution of occupation times for a class of Markov
processes, we identify the entropy production as the large deviation rate function, up to leading
order when expanding around a detailed balance dynamics. In that way, the minimum entropy
production principle is recognized as a consequence of the structure of dynamical fluctuations,
and its approximate character gets an explanation. We also discuss the subtlety emerging when
applying the principle to systems whose degrees of freedom change sign under kinematical
time-reversal.
% -----------------------------------------------------------------------------------------------------------
\normalsize
\section{Introduction}

Over the last century many attempts have been made to give a
variational characterization of nonequilibrium conditions.  The
motivation was often found in the successes of variational methods
in mechanics and in equilibrium statistical thermodynamics. Many
so called {\it ab initio} methods in solid state physics have a
variational character. For nonequilibrium purposes the best known
but also widely criticized variational principle, that of the
minimum entropy production principle (MinEP), goes back to the
work of Ilya Prigogine, \cite{P}.  In the present paper we will
restrict us to the version of the MinEP for Markov processes as
was first described and proven by Klein and Meijer for some
specific Markov models, see also~\cite{KM,ELS,MN}.\\

As has been clear since a long time, the MinEP is only valid in some approximation. Without doubt,
one restriction is that the system must be close to equilibrium, allowing only for a small
breaking of the detailed balance condition; that is often referred to as the regime of
irreversible thermodynamics. Yet, the situation is more subtle and there have appeared examples in
the literature violating the MinEP  even close to equilibrium,
\cite{L1,L2}. The situation is even more complicated and downright
controversial when dealing with examples of macroscopic physics, where both positions and
velocities mix and things appear to
depend on the level of coarse graining.\\
 At any rate and because of
the enormous advantages of variational characterizations, there has been a continued interest in
the nature of Prigogine's MinEP. It remains therefore very interesting to see if the principle can
be understood, not only by direct verification as was done in~\cite{KM,ELS,MN}, but also from the
context of fluctuation theory. After all, also in equilibrium statistical physics there is an
intimate relation between the variational principle characterizing equilibrium and the structure
of equilibrium fluctuations. The very reason why thermodynamic equilibrium is characterized by
maximum entropy or, depending on the context, by minimum Helmholtz or Gibbs free energy, is
exactly because these thermodynamic potentials also appear as rate functions in the
exponents governing equilibrium probabilities.\\

We show in this paper that a relation
 exists between the MinEP and the structure of
steady state fluctuations for Markov processes.
 Our main finding is that the entropy production
naturally emerges when analyzing the fluctuations of occupation times, first studied in the
general context of the theory of large deviations by Donsker and Varadhan~\cite{DV}. We show that
in the close-to-equilibrium regime and when the state variable is even under time-reversal, the
Donsker-Varadhan (DV-) functional coincides to leading order with the entropy production rate.
When the state variables are odd under time-reversal, such as for the electric current in the
famous counter example of \cite{L1,L2}, that affine relation between entropy production and the
DV-functional is no longer valid.  It remains of course generally true that the variational
principle associated with the DV-functional is a valid generalization of the MinEP. Yet, a useful
scheme for the computation of the DV-functional for
processes far from equilibrium remains an open problem.\\

The structure of the paper is as follows. In Section~\ref{sec: reversible} we present a brief
introduction to the large deviation theory of occupation times. In the mathematical details we
often restrict ourselves to the case of continuous time and irreducible Markov processes on a
finite state space. Many of the arguments have however a larger validity.  For example, for a
detailed balanced dynamics the DV-functional can be computed explicitly; we review that result in
Section~\ref{sec: detailed balance} with a new proof that is not restricted to finite state spaces.\\
Our main result follows from a perturbative evaluation of the DV-functional close to equilibrium
and is contained in Section~\ref{sec: DV-perturbatively}, first on a formal and general level and
then rigorously for finite state space. In Section~\ref{sec: entropy production} we explain how
and when the leading order of the DV-functional is related to the physical entropy production.
That relation is formulated in our main
Theorem~\ref{thm: main}.\\
We end with a variety of remarks and conclusions  in Section~\ref{sec: MEPP}. In particular, we
briefly explain there
the situation for Landauer's counterexample, \cite{L1,L2}.\\

%paragraph deleted.

% ------------------------------------------------------------------------------------------------------------
\section{Large deviations for the occupation times}
\label{sec: reversible}

  Suppose that $(X_t)_{t \geq 0}$ is a
stationary ergodic Markov process.  For most of what follows, we do not need to specify whether it
is a jump process or a diffusion process, and on what space. Yet, it is sufficiently instructive
and mathematically non trivial to keep in mind a Markov process on a finite space which is
irreducible.  We are interested in the fraction of time that $X_t$ spends in some set $A$ of
states. Formally, we define the empirical distribution $p_T$ as
\begin{equation}\label{timea}
  p_T(A) = \frac 1{T} \,\int_0^T\, \id t\, \de_{X_t \in A}
\end{equation}
($\de_{X_t \in A} = 1$ if $X_t \in A$ and zero otherwise.) As we assume a unique stationary
measure $\rho$, we have that almost surely $p_T(A) \rightarrow \rho(A)$ as $T\uparrow +\infty$, by
ergodicity. Yet there are fluctuations around that average and we can ask how big they are.  That
is a subject in the theory of large deviations and the answer is given by the asymptotic formula
\[
  \bsP^T[p_T \simeq \mu]  \simeq \exp[-T I(\mu)]
\]
that has to be understood in a logarithmic sense after taking the limit $T \uparrow \infty$. The
rate function $I$ has been found by Donsker and Varadhan \cite{DV,DZ,dH} in the form
\begin{equation}\label{eq: imu}
  I(\mu) = \sup_{g>0} \, -\Bigl\langle\frac{L g}{g}\Bigr\rangle_\mu
\end{equation}
where $L$ is the generator of the Markov process and
$\langle\cdot\rangle_\mu$ denotes the expectation under the
measure $\mu$. For a finite state space $\Omega$,
\[
 Lg(x) =
\sum_{y\in \Omega} k(x,y)[g(y) - g(x)],\qquad x\in \Omega
\]
where $k(x,y)\geq 0$ is the rate for the transition $x\rightarrow y$.\\
The Donsker-Varadhan (DV-)functional is always nonnegative,
$I(\mu) \geq 0$, and the equality takes place if and only if $\mu
= \rho$ is the invariant measure.  For the precise
mathematical formulation we refer to \cite{DZ,dH,DV}.\\

In general, the DV-functional~\eqref{eq: imu} is not so simple to compute explicitly, the main
problem of course being to find the maximizer $g$.  An important case where the solution has been
known and is explicit is a reversible (or detailed balance) dynamics. These basic facts are
reviewed in the next section; our formulation is slightly more general than those provided by the
standard references~\cite{DV,dH}.  The rest of the paper is then devoted to identifying the
leading term in the DV-functional for a dynamics breaking the detailed balance.

\section{Detailed balance dynamics}\label{sec: detailed balance}

Suppose that for any pair of real-valued functions $\phi$ and $\psi$,
\begin{equation}\label{eq: db-path}
  \langle \phi(x_0)\, \psi(x_\tau) \rangle_\rho =
   \langle \phi(x_\tau)\, \psi(x_0)
  \rangle_\rho
\end{equation}
where $\langle\cdot\rangle_\rho$ is the expectation under the
stationary Markov process. The corresponding symmetry of the
generator can be obtained under
\[
\lim_{\tau\downarrow 0}\frac 1{\tau}\langle \phi(x_0)\,
[\psi(x_\tau) - \psi(x_0)] \rangle_\rho = \langle \phi\,L\psi
\rangle_\rho
\]
%Taking the invariant (reversible) measure $\rho$ as a reference, to a measure $\mu$ we assign its
%density $f = \frac{\id\mu}{\id\rho}$. A basic result is the following:

\begin{theorem}\label{thm: DV-db}
Under condition \eqref{eq: db-path}, the DV-functional is
\begin{equation}\label{eq: dv-reversible}
  I(\mu) = -\bigl\langle \sqrt{f}\, L \sqrt{f} \,\bigr\rangle_\rho
\end{equation}
where $f = \frac{\id\mu}{\id\rho}$ is the density of $\mu$ with respect to the reversible measure
$\rho$.
\end{theorem}

\begin{remark}
One recognizes the Dirichlet form $\scD(g,g) = -\langle g\, L g
\rangle_\rho$ which is related to the spectral gap by
\begin{equation}
  \De = \inf_{g:\, \langle g \rangle_\rho = 0}
  \frac{\scD(g,g)}{\langle g^2 \rangle_\rho}
\end{equation}
As a consequence, one has the bound
\begin{equation}
\begin{split}
  I(\mu) &= \scD(\sqrt{f},\sqrt{f}) =
  \scD(\sqrt{f} - \langle \sqrt{f} \rangle_\rho, \sqrt{f} - \langle \sqrt{f} \rangle_\rho)
\\
  &\geq \De[\langle f \rangle_\rho - \langle \sqrt{f} \rangle_\rho^2]
  = \De[1 - \langle \sqrt{f} \rangle_\rho^2]
\end{split}
\end{equation}
\end{remark}

\begin{proof}
A standard proof for finite state space can be found e.g.\ in~\cite{dH}. Here we present a new
variant of that argument that works for a general (detailed balanced) Markov process.

From~\eqref{eq: imu},
\begin{equation}
\begin{split}
  I(\mu) &= \sup_{g > 0} \lim_{\tau \downarrow 0} \frac{1}{\tau}
  \Bigl[ 1 - \Bigl\langle
  \frac{e^{\tau L} g}{g} \Bigr\rangle_\mu \Bigr]
\\
  &= \sup_{g > 0} \lim_{\tau \downarrow 0} \frac{1}{\tau}
  \Bigl[ 1 - \Bigl\langle \frac{f(x_0)  g(x_\tau)}{g(x_0)} \Bigr\rangle_\rho \Bigr]
\end{split}
\end{equation}
Using reversibility~\eqref{eq: db-path} we subsequently get
\begin{equation}
\begin{split}
  \Bigl\langle \frac{f(x_0) g(x_\tau)}{g(x_0)}
  \Bigr\rangle_\rho &=
  \frac{1}{2} \Bigl\langle
  \frac{f(x_0) g(x_\tau)}{g(x_0)} +
  \frac{f(x_\tau) g(x_0)}{g(x_\tau)} \Bigr\rangle_\rho
\\
  &= \frac{1}{2} \Biggl\langle \Biggl( \sqrt{\frac{f(x_0)
   g(x_\tau)}{g(x_0)}}
  - \sqrt{\frac{f(x_\tau) g(x_0)}{g(x_\tau)}}\,
   \Biggr)^2 \Biggr\rangle_\rho
\\
  &\hspace{15mm} + \bigl\langle \sqrt{f(x_0) f(x_\tau)}
  \,\bigr\rangle_\rho
\\
  &\geq \langle \sqrt{f(x_0) f(x_\tau)} \rangle_\rho
\\
  &= \langle \sqrt{f} e^{\tau L} \sqrt{f} \rangle_\rho
\end{split}
\end{equation}
which is an optimal lower bound since the equality is attained if (and only if for an irreducible
dynamics) $g \propto \sqrt{f}$. Hence,
\begin{equation}
  I(\mu) = \lim_{\tau \downarrow 0} \frac{1}{\tau}
  \bigl[ 1 - \langle \sqrt{f}\, e^{\tau L} \sqrt{f} \rangle_\rho \bigr]
  = - \langle \sqrt{f}\, L \sqrt{f} \rangle_\rho
\end{equation}
as claimed.
\end{proof}

%\subsection{Finite state space}\label{sec: finite space}
%A most simple example is provided by the Markov process on the space $\Om = \{1,\ldots,N\}$, with
%the rates $\la(x,y)|_{x,y \in \Om}$, and the unique invariant measure denoted by $\rho$. The
%Kolmogorov generator is
%\begin{equation}
%  (L g)(x) = \sum_{y \neq x} \la(x,y)[g(y) - g(x)]
%\end{equation}
%and the DV-functional takes the form
%\begin{equation}\label{eq: rate function}
%  I(\mu) = \sum_{x, y \neq x} \mu(x) \la(x,y)
%  -\inf_{g > 0} \sum_{x, y\neq x} \mu(x) \la(x,y) \frac{g(y)}{g(x)}
%\end{equation}
%Under the detailed balance condition $\rho(x) \la(x,y) = \rho(y) \la(y,x)$, this further equals to
%\begin{equation}
%  I(\mu) = \sum_{x,y\neq x} \mu(x) \la(x,y) \Bigl[ 1 - \sqrt{\frac{\mu(x)\rho(y)}{\mu(y)\rho(x)}}\Bigr]
%\end{equation}
%as immediately follows from the following computation, we write again $\mu(x) = \rho(x) f(x)$:
%\begin{equation}
%\begin{split}
%  \inf_{g > 0} \sum_{x,y \neq x} &\rho(x) \la(x,y) \frac{f(x)g(y)}{g(x)}
%\\
%  &= \frac{1}{2} \inf_{g > 0} \sum_{x,y\neq x} \rho(x)\la(x,y)
%  \Bigl[\sqrt{\frac{f(x) g(y)}{g(x)}} - \sqrt{\frac{f(y) g(x)}{g(y)}} \Bigr]^2
%\\
%  &+ \sum_{x,y\neq x} \rho(x) \la(x,y) \sqrt{f(x) f(y)}
%\\
%  &= \sum_{x,y\neq x} \rho(x) \la(x,y) \sqrt{f(x) f(y)}
%\end{split}
%\end{equation}
%The above argument is standard \cite{dH}.

%One can easily check that this expression is compatible with the
%more general notation of \eqref{imu}, for
%\[
%\nu L(x) =
%\]
%\subsection{Langevin dynamics}
%The second class of dynamics is given by the Langevin equation....

\section{Perturbative evaluation of the DV-functional}\label{sec: DV-perturbatively}

%paragraph deleted

\subsection{Formal derivation}\label{sec: DV-derivation}

Fix a reference detailed balance dynamics with generator $L_0$ and with reference measure
$\rho^0$, as in Section \ref{sec: detailed balance}. For a measure $\mu$ we write $f =
\frac{\id\mu}{\id\rho^0}$ for its density with respect to
$\rho^0$. A simple computation gives
\begin{equation}
\begin{split}
  \de \Bigl\langle \frac{f}{g} L g \Bigr\rangle_{\rho^0} &=
  \Bigl\langle -\frac{f}{g^2} \de g\, L g + \frac{f}{g} L \de g\Bigr\rangle_{\rho^0}
\\
  &= \Bigl\langle \Bigl(-\frac{f}{g^2} L g + L^+ \frac{f}{g}\Bigr) \de g \Bigr\rangle_{\rho^0}
\end{split}
\end{equation}
where the adjoint $L^+$ is defined by
\begin{equation}\label{eq: adjoint}
  \langle \phi\, L \psi \rangle_{\rho^0} = \langle \psi\, L^+ \phi \rangle_{\rho^0}
\end{equation}
on real functions. Hence, searching for the maximizer $g^*$ of \eqref{eq: imu} normalized to
$\langle g^* \rangle_{\rho^0} = 1$, we need to solve the equation
\begin{equation}\label{eq: minimizer}
  \frac{f}{g^{*2}} L g^* = L^+ \frac{f}{g^*}
\end{equation}
Note that for $L= L_0 = L_0^+$ that equation has a solution
$g^* = \sqrt{f}/\langle \sqrt{f} \rangle_{\rho^0}$, in agreement with the conclusions of
Section~\ref{sec: detailed balance}.\\
Next, for a close to equilibrium dynamics and for small fluctuations we expand $L$,
$f$, and $g$ in power series,
\begin{align}\label{eq: exp1}
  L^\ep &= L_0 + \ep L_1 + \ep^2 L_2 + \ldots
\\\label{eq: exp2}
  f^\ep &= 1 + \ep f_1 + \ep^2 f_2 + \ldots
\\
  g^\ep &= 1 + \ep g_1 + \ep^2 g_2 + \ldots
\end{align}
and solve~\eqref{eq: minimizer} perturbatively. Up to order
$\ep$ it yields
\begin{equation}\label{eq: g1}
  2 L_0 g_1^* = L_0 f_1 + L_1^+ 1
\end{equation}
which is to be solved under the normalization constraint $\langle g_1^* \rangle_{\rho^0} = 0$. That
can be achieved as follows. Writing $\frac{\id\rho^\ep}{\id\rho^0} = h^\ep$ for the density of the
(presumably unique for small $\ep$) stationary measure under $L^\ep$ with respect to the reference
reversible measure, the stationary equation $\rho^\ep L^\ep = 0$ can be equivalently written as
$(L^{\ep})^+ h^\ep = 0$. Expanding again
$h^\ep = 1 + \ep h_1 + \ldots$, we find that
$h_1$ verifies
\begin{equation}
  L_0 h_1 = -L_1^+ 1
\end{equation}
and, by definition, $\langle h_1 \rangle_{\rho^0} = 0$. As a consequence, $g_1^* = (f_1 - h_1)/2$
is a solution of~\eqref{eq: g1}. Provided that
$g^*$ is in fact a global maximum, the DV-functional~\eqref{eq: imu} becomes, up to leading order,
\begin{equation}\label{eq: DV-explicit}
\begin{split}
  I^\ep(\mu^\ep) &= -\frac{\ep^2}{4} \langle f_1 L_0 f_1 - h_1 L_0 h_1 + 2 L_1 f_1 - 2 L_1 h_1
  \rangle_{\rho^0} + o(\ep^2)
\\
  &= -\Bigl\langle \sqrt{\frac{f^\ep}{h^\ep}} L^\ep \sqrt{\frac{f^\ep}{h^\ep}}\, \Bigr\rangle_{\rho^0}
  + o(\ep^2)
\end{split}
\end{equation}
The functional $I^\ep$ itself obviously also depends on $\ep$ as from \eqref{eq: exp1}; we are
dealing with a dynamics close to a reference reversible dynamics. Observe
 that, since $f^\ep\,\id\rho^\ep =
h^\ep\,\id\mu^\ep$, the leading term in the DV-functional \eqref{eq: DV-explicit} (always for small
deviations from equilibrium) resembles the DV-functional~\eqref{eq: dv-reversible} for the case of
detailed balance.  In \eqref{eq: DV-explicit} that leading term is now of order $\ep^2$.

\subsection{Rigorous result}\label{sec: chains}

The above formal perturbative argument can be justified on a mathematically precise level. In the
present section we refine the above reasoning by restricting ourselves to the framework of
continuous time Markov dynamics with a finite state space. Note that many of the standard
nonequilibrium examples of stochastic lattice gases or interacting particle systems on a finite
graph are thus included~\cite{ELS,L,S}. Observe also that some precision or justification is
indeed needed, as one can otherwise construct counter examples to the results that will follow.
Other ``infinite'' or ``continuous'' models including diffusion processes, still require
additional estimates for a proper mathematical treatment, that we are not giving here though; we
will comment on one important example in Section \ref{sec: MEPP}. On the whole and perhaps
surprisingly, even only to first order around equilibrium, a general and mathematically precise
identification of the DV-functional does not appear easy.\\

We fix a finite state space $\Omega$, which will serve as vertex set for irreducible directed
graphs respectively with rates $k^0(x,y)\geq 0$ (reference detailed balance) and with rates
$k^\ep(x,y)\geq 0$ (perturbation) between the states $x\rightarrow y$.  We assume the reference rates
$k^0(x,y)$ define an ergodic Markov process with the stationary distribution
$\rho^0 > 0$ and such that
$\rho^0(x) k^0(x,y) = \rho^0(y) k^0(y,x)$, sufficient for the reversibility in \eqref{eq: db-path}.
The perturbed rates $k^\ep(x,y)$ defined for $|\ep| \leq \ep_0$ with some $\ep_0>0$, are assumed to
be a smooth modification of the $k^0(x,y)$.  For small enough
$\ep$ the perturbed dynamics is hence ergodic too, with a unique invariant distribution $\rho^\ep > 0$
which is a smooth modification of $\rho^0$.
%We write $h^\ep(x) = \rho^\ep(x)/\rho^0(x)$.

The modified dynamics has the generator
\begin{equation}
  L^\ep g(x) = \sum_{x,y \neq x} k^\ep(x,y) [g(y) - g(x)]
\end{equation}

%For a continuous time Markov chain the states $x,y,\ldots$ are elements of a set $\Om_N =
%\{1,2,\ldots,N\}$ and the Kolmogorov generator has the general form
%\begin{equation}
%  L^\ep f(x) = \sum_{x,y\neq x} k^\ep(x,y) [f(y) - f(x)]
%\end{equation}
%with transition rates $k^\ep(x,y) \geq 0$, $x \neq y \in \Om_N$, $|\ep| \leq \ep_0$. It is assumed
%to be a continuous modification of a fixed ergodic detailed balanced Markov chain with rates
%$k^0(x,y)$ and a reversible measure $\rho^0$, both related by
%$\rho^0(x) k^0(x,y) = \rho^0(y) k^0(y,x)$, see \eqref{eq: db}-\eqref{eq: db-path}.

We further denote
\begin{align}
\nonumber
  M_{+1} &= \{g>0;\,\langle g \rangle_{\rho^0} = 1\}\,, \qquad
  M_{+1}^\de = \{g \in M_{+1};\,g(x) \geq \de, x\in \Omega\}
\\\intertext{and we consider the functional}
\label{eq: J}
  J^\ep_f(g) &=
\sum_{x,y \neq x} \rho^0(x) f(x) k^\ep(x,y) \Bigl[ 1 - \frac{g(y)}{g(x)}
\Bigr]
\end{align}
for $f\in M_{+1}^\de$ on $g\in   M_{+1}$.

% ------------------------------------------------------------------------------------- THEOREM
\begin{proposition}\label{prop}
Suppose that  $f  \in M_{+1}^\de$ for some $\de
> 0$. For all sufficiently small $|\ep|$, the functional
$J^\ep_f$ has a unique maximizer $g^{*\ep}(f)$ in $M_{+1}$, and
$g^{*\ep}(f) \stackrel{\ep \downarrow 0}{\rightarrow} \sqrt{f} /
\langle \sqrt{f} \rangle_{\rho^0}$, uniformly in $ M_{+1}^\de$.
\end{proposition}

\begin{theorem}\label{thm: finite space}
If $\mu^\ep$ is a smooth deformation of $\mu^0 = \rho^0$, then the DV-functional $I^\ep(\mu^\ep)$
has a Taylor expansion in $\ep$ around $\ep=0$, with leading term
\begin{equation}\label{eq: DV-explicit2}
  I^\ep(\mu^\ep) = -\Bigl\langle \sqrt{\frac{\id\mu^\ep}{\id\rho^\ep}}\,
  L^\ep \sqrt{\frac{\id\mu^\ep}{\id\rho^\ep}}\,
  \Bigr\rangle_{\rho^0} + o(\ep^2)
\end{equation}
\end{theorem}
The proofs are postponed to Section~\ref{sec: finite space - proof}.

\section{Relation with entropy production}\label{sec: entropy production}
We proceed with the \emph{physical} interpretation of the formula~\eqref{eq: DV-explicit2} for the
DV-functional. It will turn out that~\eqref{eq: DV-explicit2} equals the excess of entropy
production with respect to the stationary entropy production.  Clearly, to explain, we need some
physical context for the dynamics itself. However in order to avoid relying solely on concrete
examples, we can start from the quite general observation that the physical entropy production as
a variable on path-space is measuring the breaking of time-reversal symmetry. That has been argued
for at various places, see e.g.~\cite{MN,M} and references therein.  When the distribution at time
zero is given by $\mu$, then the entropy production over the time interval $[0,\tau]$, is just the
relative entropy of the path-space distribution $\bsP^\tau_{\mu}$ with respect to its
time-reversal:
\begin{equation}\label{eq: even}
  \dot S^\tau(\mu) =
  \Bigl\langle \log\frac{\id\bsP^\tau_{\mu}}{\id\bsP^\tau_{\mu_\tau}\Th}
  \Bigr\rangle_{\mu}
\end{equation}
where $(\Th \om)_t = \om_{\tau - t}$ is the time reversal of the trajectory $\om$ and $\mu_\tau$
is the evolved distribution at time $\tau$, i.e., the solution of the Master equation
$\frac{\id\mu_t}{\id t} = \mu_t L$, $\mu_0 = \mu$. Since the
process is Markovian, the mean entropy production can be written as $\dot S^\tau(\mu) =
\int_0^\tau \si(\mu_t)\,\id t$ where
\begin{equation}\label{eq: even-rate}
  \si(\mu) = \lim_{\tau \downarrow 0} \frac{\dot S^\tau(\mu)}{\tau}
\end{equation}
is the mean entropy production rate. Taken as a functional on distributions $\mu$, \eqref{eq:
even-rate} is the crucial quantity to be discussed in the present section.  In particular we can
evaluate it under the same conditions as for Theorem \ref{thm: finite space}.  It means that we
evaluate the entropy production rate in $\mu^\ep$ and that we have a dynamics that is close to
equilibrium, indicated by changing the notation $\sigma$ to
$\sigma^\ep$.  The main result of the paper is then summarized in
the following general and remarkable relation:
\begin{theorem}\label{thm: main}
Under the conditions of Theorem  \ref{thm: finite space},
\[
I^\ep(\mu^\ep) = \frac{1}{4}[\si^\ep(\mu^\ep) - \si^\ep(\rho^\ep)] + o(\ep^2)
\]
\end{theorem}

Before we give the proof of that Theorem, we briefly remind the reader of the physical context of
entropy production, at least within the limited set-up of Markov jump processes.  We refer to
\cite{QQT,JMP,ELS,MN,KM} for additional material.

\subsection{Entropy production in Markov jump processes}
For the Markov jump processes of Section~\ref{sec: chains} the entropy production rate~\eqref{eq:
even-rate} becomes
\begin{equation}\label{eq: sigma-chain}
  \si(\mu) = \sum_{x,y \neq x} \mu(x) k(x,y) \log \frac{\mu(x) k(x,y)}{\mu(y) k(y,x)}
\end{equation}
In the case of detailed balance, $\rho(x)k(x,y)=\rho(y) k(y,x)$, it is easily verified that
$\si(\mu)$ is the time derivative of the relative entropy:
%see e.g.~\cite{brus}:
\begin{equation}\label{eq: ep-db}
\begin{split}
  \si(\mu) &= \sum_x \log \frac{\mu(x)}{\rho(x)} \sum_{y \neq x}
  \bigl[ \mu(x) k(x,y) - \mu(y) \rho(y,x) \bigr]
\\
  &= -\sum_x \log \frac{\mu(x)}{\rho(x)}\, \frac{\id\mu_t(x)}{\id t} \Bigl|_{t=0}
\\
  &= -\frac{\id}{\id t}S(\mu_t \rel \rho)|_{t=0},\qquad S(\mu \rel \rho) =
  \sum_{x} \mu(x) \log \frac{\mu(x)}{\rho(x)}
\end{split}
\end{equation}
When there is a driving away from equilibrium, there is  some mean entropy production even in the
stationary regime.  To be specific, assume that each state $x$ is given an energy $E(x)$ and that
the transition $x \leftrightarrow y$ is possible thanks to the interaction with a heat reservoir
at inverse temperature $\be(x,y) = \be(y,x)$. The rates are taken to satisfy the local detailed
balance condition
\begin{equation}
  \frac{k(x,y)}{k(y,x)} = e^{\be(x,y) [E(x) - E(y)]}
\end{equation}
For a motivation, see~\cite{ELS,MN}. As a reference we have the Boltzmann-Gibbs distribution
$\rho(x) \propto e^{-\be E(x)}$ with
$\be$ some reference inverse temperature.

Entropy production rate~\eqref{eq: sigma-chain} can  be split into a contribution which is
associated to the system and can be written as the time derivative of some entropy function, and a
part measuring the change of entropy in the environment, i.e.,
\[
\si(\mu) = \si_S(\mu) + \si_R(\mu)
\]
For the system part we take, compare with~\eqref{eq: ep-db},
\begin{equation}\label{eq: ep-system}
\begin{split}
  \si_S(\mu) &= \sum_{x,y \neq x} \mu(x) k(x,y) \log \frac{\mu(x) \rho(y)}{\mu(y) \rho(x)}
\\
  &= -\frac{\id}{\id t}S(\mu_t \rel \rho)|_{t=0}
%  &= \sum_{x} [\log \mu(x) + \be E(x)]
%  \sum_{y \neq x} [\mu(x) k(x,y) - \mu(y) k(y,x)]
%\\
%  &= -\sum_x [\log \mu(x) + \be E(x)] \frac{\id \mu_t(x)}{\id t}
\\
  &= \frac{\id}{\id t} \bigl[ S(\mu_t) - \be \langle E \rangle_{\mu_t} \bigr] \bigl|_{t=0}
\end{split}
\end{equation}
with $S(\mu) = -\sum_x \mu(x) \log \mu(x)$ the Shannon entropy and
$\langle E \rangle_\mu = \sum_x \mu(x) E(x)$ the mean energy. Hence, $\si_S(\mu)$ is
recognized as ($-\be$ times) the rate of change in the free energy.
%One can read that as {\bf CHECK!}
%$\be \times \text{\emph{ (the rate of free energy decrease)}}$.

The environment part is then
%\begin{equation}
 % \si_R(\mu) = \sum_{x,y \neq x} \mu(x) k(x,y) \log \frac{\rho^0(x) k(x,y)}{\rho^0(y) k(y,x)}
%\end{equation}
%Let us explicitly check that its interpretation is correct
%provided the model is constructed within a meaningful physical
%context. A
\begin{equation}
\begin{split}
  \si_R(\mu) &=\sum_{x,y \neq x} \mu(x) k(x,y) \log \frac{\rho(x) k(x,y)}{\rho(y) k(y,x)}
  \\
  &= \frac{1}{2}
  \sum_{x,y \neq x} \bigl[ \be(x,y) - \be \bigr]\, \bigl[ E(x) - E(y)\bigr]\,
  \bigl[\mu(x) k(x,y) - \mu(y) k(y,x) \bigr]
\\
  &=\frac{1}{2} \sum_{x,y \neq x} \bigl[ \be(x,y) - \be \bigr] \langle J_E(x,y) \rangle_\mu
\end{split}
\end{equation}
where $\langle J_E(x,y) \rangle_\mu = \langle J_E(y,x)
\rangle_\mu$ is the mean energy transfer, or heat, to the
reservoir associated with the transitions $x \leftrightarrow y$. In other words, $\si_R(\mu)$ is
the change of entropy in the environment plus the term
%{\bf CHECK!} $-\be \times
%\text{ \emph{(the rate at which the energy of the system decreases)}}$.
\[
  \be \sum_x E(x) \bigl[\mu(x) k(x,y) - \mu(y) k(y,x) \bigr]
  = \be \frac{\id}{\id t} \langle E \rangle_{\mu_t} |_{t=0}
\]
which is just the counter term we have subtracted from the system
part \eqref{eq: ep-system}.
%paragraph deleted.
\subsection{Proof of Theorem~\ref{thm: main}}\label{sec: entropy-per}

Following our general strategy, we compute \eqref{eq: even-rate} by a  perturbation expansion
around a reference detailed balanced dynamics.  Again, the expansion is mathematically fully
justified for a finite state space, at least under the conditions of the
theorem.\\

We split the entropy production rate similarly as in the previous
section, taking now the invariant distribution $\rho^0$
corresponding to $\ep = 0$ as the reference: starting from
\eqref{eq: even-rate},
\begin{equation}\label{eq: sigma}
\begin{split}
  \si(\mu) &= \lim_{\tau \downarrow 0}\frac{1}{\tau}
  \Bigl\langle
  \log\frac{\id\mu}{\id\rho^0}(\om_0) - \log\frac{\id\mu_\tau}{\id\rho^0}(\om_\tau)
%  -\log\frac{\id\mu_t}{\id\mu}(\om_\tau)
  + \log\frac{\id\bsP^\tau_{\rho^0}}{\id\bsP^\tau_{\rho^0}\Th} \Bigr\rangle_\mu
\\
%  &= -\Bigl\langle L \log\frac{\id\mu}{\id\rho^0} \Bigr\rangle_\mu
%  - \lim_{\tau\downarrow 0} \frac{1}{\tau}
%  \Bigl\langle \log\frac{\id\mu_t}{\id\mu} \Bigr\rangle_{\mu_t}
  &= \lim_{\tau \downarrow 0} \frac{1}{\tau} \Bigl[
  \Bigl\langle \log \frac{\id\mu}{\id\rho^0} \Bigr\rangle_\mu
  - \Bigl\langle \log \frac{\id\mu_\tau}{\id\rho^0} \Bigr\rangle_{\mu_\tau} \Bigr]
  + \lim_{\tau\downarrow 0} \frac{1}{\tau}
  \Bigl\langle \log\frac{\id\bsP^\tau_{\rho^0}}{\id\bsP^\tau_{\rho^0}\Th} \Bigr\rangle_\mu
\end{split}
\end{equation}
The first term is the limit
\begin{equation}\label{eq: EP-system}
\begin{split}
  \si_S(\mu) &= \lim_{\tau \downarrow 0} \frac{1}{\tau}
  \Bigl[ \Bigl\langle \log \frac{\id\mu}{\id\rho^0} \Bigr\rangle_\mu
  - \Bigl\langle \log \frac{\id\mu}{\id\rho^0} \Bigr\rangle_{\mu_\tau}
  - \Bigl\langle \frac{\id\mu_\tau}{\id\mu}
  \log \frac{\id\mu_\tau}{\id\mu} \Bigr\rangle_{\mu} \Bigr]
\\
  &= -\langle L \log f \rangle_\mu
  -\Bigl\langle \frac{\id\mu L}{\id\mu} \Bigr\rangle_\mu
\\
  &= -\langle L \log f \rangle_\mu
\end{split}
\end{equation}
Expanding both $\mu \equiv \mu^\ep$ and $L \equiv L^\ep$ as
in~\eqref{eq: exp1}--\eqref{eq: exp2}, we get
\begin{equation}
  \si^\ep_S(\mu^\ep) = -\langle f^\ep L^\ep \log f^\ep \rangle_{\rho^0}
  = -\ep^2 \langle f_1 L_0 f_1 + L_1 f_1 \rangle_{\rho^0} + o(\ep^2)
\end{equation}
Similarly, for the second term in~\eqref{eq: sigma}, now  denoted by $\si_R^\ep$, we have
\begin{equation}
\begin{split}
  \si^\ep_R(\mu^\ep) &= \lim_{\tau \downarrow 0} \frac{1}{\tau}
  \Bigl\langle \log\frac{\id\bsP^{\tau,\ep}_{\rho^0}}{\id\bsP^{\tau,0}_{\rho^0}}(\om)
  - \log\frac{\id\bsP^{\tau,\ep}_{\rho^0}}{\id\bsP^{\tau,0}_{\rho^0}}(\th\om)
  \Bigr\rangle_{\mu^\ep}^{\ep}
\\
  &= \lim_{\tau \downarrow 0} \frac{1}{\tau}
  \Bigl\langle \frac{\id\mu^\ep}{\id\rho^0}(\om_0)
  \frac{\id\bsP^{\tau,\ep}_{\rho^0}}{\id\bsP^{\tau,0}_{\rho^0}}(\om)
  \Bigl\{ \frac{\id\bsP^{\tau,\ep}_{\rho^0}}{\id\bsP^{\tau,0}_{\rho^0}}(\om)
  - \frac{\id\bsP^{\tau,\ep}_{\rho^0}}{\id\bsP^{\tau,0}_{\rho^0}}(\Th\om)
  \Bigr\} \Bigr\rangle_{\rho^0}^{0} + o(\ep^2)
\\
  &= \lim_{\tau \downarrow 0} \frac{1}{\tau}
  \Bigl\langle
  \frac{\id\bsP^{\tau,\ep}_{\rho^0}}{\id\bsP^{\tau,0}_{\rho^0}}(\om)
  \Bigl\{ \frac{\id\mu^\ep}{\id\rho^0}(\om_0) - \frac{\id\mu^\ep}{\id\rho^0}(\om_\tau)
  \Bigr\} \Bigr\rangle_{\rho^0}^{0}
\\
  &+ \lim_{\tau \downarrow 0} \frac{1}{2\tau}
  \Bigl\langle \Bigl\{
  \frac{\id\bsP^{\tau,\ep}_{\rho^0}}{\id\bsP^{\tau,0}_{\rho^0}}(\om)
  - \frac{\id\bsP^{\tau,\ep}_{\rho^0}}{\id\bsP^{\tau,0}_{\rho^0}}(\Th\om)
  \Bigr\}^2 \Bigr\rangle_{\rho^0}^{0} + o(\ep^2)
\\
  &= -\langle L^\ep f^\ep \rangle_{\rho^0} + \De^\ep + o(\ep^2)
\\
  &= -\ep^2 \langle L_1 f_1 \rangle_{\rho^0} + \De^\ep + o(\ep^2)
\end{split}
\end{equation}
where $\bsP^{\tau,0}_{\rho^0}$ and $\langle \cdot \rangle_{\rho^0}^{0}$ refer to the path-space
distribution  under the reference detailed balance dynamics ($\ep=0$) started from $\rho^0$.  The
term
$\De^\ep$ is simply independent of $f^\ep$. All in all we have found, up to leading order,
\begin{equation}
  \si^\ep(\mu^\ep) = -\ep^2 \langle f_1 L_0 f_1 + 2 L_1 f_1 \rangle_{\rho^0} + \De^\ep + o(\ep^2)
\end{equation}
Comparing with the result~\eqref{eq: DV-explicit} or \eqref{eq: DV-explicit2} finishes the proof.

\section{Proof of Proposition~\ref{prop} and of
Theorem~\ref{thm: finite space}}\label{sec: finite space - proof}

 Let $0 < \de < 1$ be given and fix  $\mu$ by giving $f
= \frac{\id\mu}{\id\rho^0} \in M_{+1}^\de$. In order to localize the maximizer $g^{*\ep}(f)$ of
the functional $J^\ep_{f}$, we decompose the set $M_{+1}$ as follows. Given $\al,\be > 0$ such
that $\al + \be < \sqrt{\de}$ we introduce
\begin{equation}
  N_{+1}^\al(\mu) = \Bigl\{g \in M_{+1};\,
  \Bigl|g(x) - \frac{\sqrt{f(x)}}{\langle \sqrt{f}\rangle_{\rho^0}}\Bigr| < \al,\,
   x\in \Omega\Bigr\}
\end{equation}
Obviously, $N_{+1}^\al(\mu) \subset M_{+1}^\be$, and writing
$[M_{+1}^\be]^c = M_{+1} \setminus M_{+1}^\be$ we have the
disjoint decomposition
\begin{equation}
  M_{+1} = N_{+1}^\al(\mu) \cup [M_{+1}^\be \setminus N_{+1}^\al(\mu)]
  \cup [M_{+1}^\be]^c
\end{equation}
In what follows we are going to prove that, choosing $\ep,
\al,\be$ small enough, the functional $J^\ep_{f}$ takes its
maximum inside $N_{+1}^\al(\mu)$, and that is unique
 by a local convexity argument.\\

We start with a lemma that follows immediately from the
assumptions. Recall that the state space is assumed finite; let
$|\Omega|=N$.
\begin{lemma}\label{lem: graph}
There is an irreducible graph $G$ with vertex set $\Omega$ and for which over all edges $(x,y)$,
$\rho^0(x) k^0(x,y) \geq \gamma$ for some $\gamma>0$.  Moreover, $k^\ep(x,y) \geq \frac{1}{2}
k^0(x,y)$ for all sufficiently small $|\ep| > 0$.
\end{lemma}

% ------------------------------------------------------------------------------- LEMMA
The next lemma states that when $g$ is outside $M_{+1}^\be$, then
$J_f^\ep(g)$ can be made very negative ($\beta \downarrow 0$).

\begin{lemma}\label{lem: tail}
For all sufficiently small $|\ep|$ and for all $g \in [M_{+1}^\be]^c$,
\begin{equation}\label{eq: J-ub}
  J_f^\ep(g) \leq C - \frac{1}{2} \ga \de \be^{-\frac{1}{N-1}}
\end{equation}
with $C$ a constant independent of $f$, $g$ and $\ep$.
\end{lemma}
\begin{proof}
From the previous lemma,
\begin{equation}
\begin{split}
  J^\ep_f(g) &= \sum_{x,y \neq x} \rho^0(x) f(x) k^\ep(x,y) \Bigl[ 1 - \frac{g(y)}{g(x)} \Bigr]
\\
  &\leq \max_x \sum_{y \neq x} k^\ep(x,y)
  -\frac{1}{2} \sum_{(x,y) \in G}
  \rho^0(x) k^0(x,y) \Bigl[\frac{f(x) g(y)}{g(x)} + \frac{f(y) g(x)}{g(y)}\Bigr]
\\
  &\leq C - \frac{1}{2} \ga \de \sum_{(x,y) \in G} \Bigl[\frac{g(y)}{g(x)} + \frac{g(x)}{g(y)}\Bigr]
\end{split}
\end{equation}
for a suitable $C$ and $\ep$ small enough.  Since $g \in
[M_{+1}^\be]^c$, there is $\bar x \in \Omega$ such that $g(\bar x)
< \be$. Hence there exists a pair $(x,y) \in G$ such that either
$g(y) \geq \be^{-\frac{1}{N-1}} g(x)$ or $g(x) \geq
\be^{-\frac{1}{N-1}} g(y)$. To see that, assume this is not true
and denote by $l(x)$ the length of the shortest path in $G$
connecting $\bar x$ and $x$. Then, using $l(x) \leq N-1$,
\begin{equation}
  \langle g \rangle_{\rho^0} \leq \max_x g(x)
  \leq g(\bar x) \max_x \be^{-\frac{l(x)}{N-1}}
  \leq \be^{-1} g(\bar x) < 1
\end{equation}
which is a contradiction.
\end{proof}
% ---------------------------------------------------------------------------- LEMMA
Now comes the statement that the maximum is also outside
$M_{+1}^\be \setminus N_{+1}^\al(\mu)$.\\
Use the shorthand $g_0 = \sqrt{f} / \langle \sqrt{f}
\rangle_{\rho^0}$.
\begin{lemma}\label{lem: continuous}
For all sufficiently small $|\ep| > 0$ we have $J^\ep_f(g) < J^\ep_f(g_0)$  whenever $g \in
M^\be_{+1}
\setminus N^\al_{+1}(\mu)$.
\end{lemma}
\begin{proof}
As clear from the proof of Theorem~\ref{thm: DV-db}, $g_0$ is the unique maximizer of $ J^0_f$ in
$M_{+1}$ due to the irreducibility assumption. Using that $M^\be_{+1}
\setminus N^\al_{+1}(\mu)$ is a compact set (in the Euclidean
metric, say),
\begin{equation}\label{eq: J0tech}
  \sup_{g \in M^\be_{+1} \setminus N^\al_{+1}(\mu)} J^0_f(g)
  < J^0_f(g_0)
\end{equation}
By the continuity of $J^\ep_f(g)$ at $\ep = 0$ which is uniform in
$g \in M^\be_{+1}$, we can choose $|\ep| > 0$ sufficiently small
so that~\eqref{eq: J0tech} extends to
\begin{equation}
  \sup_{g \in M^\be_{+1} \setminus N^\al_{+1}(\mu)} J^\ep_f(g)
  < J^\ep_f(g_0)
\end{equation}
Hence, the lemma follows.
\end{proof}
% ------------------------------------------------------------------------- LEMMA
\begin{lemma}\label{lem: concave}
%We assume $k^\ep(x,y)$, $x \neq y \in \Om$, to be twice differentiable on a neighborhood of
%$\ep = 0$.
There is $\al > 0$ such that for all sufficiently small $|\ep| > 0$,
$J^\ep_f$ is a strictly concave function in $N^\al_{+1}(\mu)$.
\end{lemma}
\begin{proof}
For any $\psi: \Om \to \bbR$, a direct computation yields
\begin{equation}\label{eq: convex}
\begin{split}
  \frac{\id^2}{\id t^2}\Bigl|_{t=0} J^0_f(g_0 + t\psi) &=
  -\sum_{x,y \neq x} \rho^0(x) k^0(x,y)
\\
  &\hspace{5mm}\times \Bigl[ \psi(x) \Bigl( \frac{f(y)}{f(x)} \Bigr)^{\frac{1}{4}}
  - \psi(y) \Bigl( \frac{f(x)}{f(y)} \Bigr)^{\frac{1}{4}} \Bigr]^2
\end{split}
\end{equation}

Since the term in the square bracket is strictly positive unless
$\psi(x)/\psi(y) = \sqrt{f(x)/f(y)}$, and using Lemma \ref{lem:
graph}, the right-hand side in~\eqref{eq: convex} is strictly negative unless $\psi \propto
\sqrt{f}$. In particular, it implies
$\id^2 J^0_f(g_0 + t\psi) / \id t^2 (t=0)$ is a strictly negative
quadratic form on the linear subspace defined by $\langle \psi
\rangle_{\rho^0} = 0$.

By continuity, it first extends to the strict negativity of the quadratic form $\id^2  J^0_f(g +
t\psi) / \id t^2 (t=0)$ on the same linear subspace and for all $g \in N_{+1}^\al(\mu)$ with some
$\al > 0$. Finally, it implies the strict negativity of $\id^2
J^\ep_f(g + t\psi) / \id t^2 (t=0)$ on $\langle \psi
\rangle_{\rho^0} = 0$ for all $g \in N_{+1}^\al(\mu)$ and for all
sufficiently small $|\ep| > 0$.
\end{proof}

\begin{proof}[Proof of Theorem~\ref{thm: finite space}]
Pick some $\al > 0$ such that Lemma~\ref{lem: concave} holds, and fix $\be > 0$ to satisfy
$J^\ep_\mu(g_0) > C - \frac{1}{2} \ga \de \be^{-\frac{1}{N-1}}$ for
all $|\ep| > 0$ small enough. Then by Lemmata~\ref{lem: tail}-\ref{lem: continuous}, the maximizer
of $J^\ep_\mu$ exists and is localized in $N_{+1}^\al$; moreover it is unique by Lemma~\ref{lem:
concave}. As the $\al$ can be chosen arbitrarily small
 (and observe that $\alpha \downarrow 0$ drives $\ep
\downarrow 0$), one also has $g^{*\ep}(\mu) \stackrel{\ep \downarrow
0}{\rightarrow} g_0$.

If $k^\ep(x,y)$, $x \neq y \in \Om$, are all differentiable then
the maximizer $g^{*\ep}$ coincides with a solution of~\eqref{eq:
minimizer} in the domain $N_{+1}^\al$; recall the latter
necessarily exists and is unique. The perturbative calculation in
Section~\ref{sec: DV-derivation} then follows by an application of
the inverse mapping theorem.
\end{proof}

\section{Conclusions: minimum entropy production
principle}\label{sec: MEPP}

\subsection{Summary}
Our analysis goes beyond merely checking MinEP; rather, it enables
to view it as a consequence of a dynamical variant of the
Einstein's formula for equilibrium fluctuations. In simple terms,
Theorem~\eqref{thm: main} reads that the probability for the
empirical distribution $p_{T}$ to coincide with some $\mu^\ep =
\rho^\ep + O(\ep)$, has the following generic structure:
\begin{equation}
  \bsP^{T,\ep}(p_{T} \simeq \mu^\ep) \propto e^{-\frac{T}{4}[\si^\ep(\mu^\ep) + o(\ep^2)]
   + o(T)}
\end{equation}
By ergodicity, the maximal probability is obtained for $p_{T} =
\rho^\ep$. According to the above it is also obtained by
minimizing the entropy production. Hence, the minimum entropy production principle emerges as an
immediate consequence of the structure of dynamical fluctuations. Moreover, its approximate status
is also understood since the relation between the entropy production and the true DV-functional is
restricted to the leading order of expansion around equilibrium. A systematic perturbation
expansion of $I^\ep(\mu^\ep)$ would provide corrections to that principle; we will not discuss
that issue now.  Some further remarks end the paper:
\subsection{Remarks}
\begin{enumerate}
\item What has been said so far about the entropy production is
subject to one further \emph{physical} condition: that the Markov
process describes the dynamics of time-reversal symmetric
variables. Only then are~\eqref{eq: even-rate} or~\eqref{eq:
sigma-chain} correct expressions for the entropy production rate.
Yet, certain observables like e.g.\ momentum or magnetic field
have the property that even a closed system dynamics cannot be
expected detailed balanced in the sense of~\eqref{eq: db-path}.
Instead, a symmetry under time-reversal can only be seen when also
the sign of these, so called time-reversal odd observables is
changed. A deeper reason why such a generalization is needed is
that the fundamental equations of motion are often second order in
time. For processes on variables that are odd under time-reversal
the above analysis needs a modification (see also the next
remark).

\item We give an example of a Gaussian Markov diffusion process
$(X_t)$. Suppose a Langevin dynamics of the form
\begin{equation}\label{rllang}
  \id X_t = (\caE - \gamma X_t)\,\id t +
   \sqrt\frac{2 \gamma}{\be }\,\id W_t
\end{equation}
with standard Wiener process $W_t$. The force $\caE$ is constant
and $\gamma>0$ is some friction coefficient.  For scalar $X_t
\in\bbR$ the process is detailed balanced in the sense
of~\eqref{eq: db-path} with respect to $\rho(\id x) \propto \exp[-
\frac{\beta}{2} (x - \frac{\caE}{\gamma})^2]\,\id x$, a Gibbs
distribution for inverse temperature $\beta$. From
Theorem~\ref{thm: DV-db} one easily computes the corresponding
DV-functional to be
\begin{equation}\label{eq: diffusion-I}
  I(\mu) = \frac{\ga}{4\be} \Bigl\langle \frac{(f')^2}{f} \Bigr\rangle_\rho\,,
  \qquad f = \frac{\id\mu}{\id\rho}
\end{equation}
Is that equal to the entropy production?  It now depends on whether $X_t$ is even or odd under
time-reversal.

Assume first that $X_t$ models the position of an overdamped oscillator. That is an even variable
and the detailed balance~\eqref{eq: db-path} is verified; the stationary process is in
equilibrium. The entropy production is found most easily from~\eqref{eq: ep-db}:
\begin{equation}\label{eq: diffusion-even}
\begin{split}
  \si(\mu) &= -\frac{\id}{\id t} S(\mu_t \rel \rho) |_{t = 0}
  = \frac{\ga}{\be} \Bigl\langle \frac{(f')^2}{f} \Bigr\rangle_\rho
\end{split}
\end{equation}
Hence, we get $\si(\mu) = I(\mu)/4$, consistent with our general
result.\\
Alternatively, suppose now that $X_t \equiv V_t$ is instead the
fluctuating velocity of a Langevin particle dragged by force
$\caE$. Although~\eqref{eq: db-path} remains valid, it no longer
expresses time-reversal invariance since the kinematical
time-reversal (changing the sign of the velocity) is not applied.
Furthermore, if $\caE \neq 0$, then $\langle \phi(v_0)\,
\psi(v_\tau) \rangle_\rho \neq \langle \phi(-v_\tau)\,
\psi(-v_0)\rangle_\rho$ breaking even a (generalized)
reversibility. In particular, there is for $\caE \neq 0$ a nonzero
stationary entropy production. That mean entropy production can be
obtained by the methods of Reference~\cite{HCN} in the form
\begin{equation}\label{eq: diffusion-odd}
  \si(\mu) = \frac{\ga}{\be} \Bigl\langle \frac{1}{f} \Bigl(f' + \frac{\be\caE}{\ga} f\Bigr)^2 \Bigr\rangle_\rho
\end{equation}
and is different from~\eqref{eq: diffusion-even}. Using that
$\langle f' \rangle_\rho = \be \langle v - \frac{\caE}{\ga}
\rangle_\mu$ and $\si(\rho) = \be \caE^2/\ga$, we obtain the
following modification of Theorem~\ref{thm: main}:
\begin{equation}
  I(\mu) = \frac{1}{4} \bigl[ \si(\mu) + \si(\rho) - 2\be\caE \langle v \rangle_\mu \bigr]
\end{equation}
In particular, the stationary distribution is now found as a
minimizer of the functional $\si(\mu) - 2\be\caE \langle v
\rangle_\mu$. Equivalently, since $\si(\rho) = \be\caE \langle v
\rangle_\rho$, the stationary measure is now characterized by a
(constrained) \emph{maximum} entropy production principle:
\begin{equation}
  \max_\mu \{\si(\mu) \rel \si(\mu) = \be\caE \langle v \rangle_\mu \} = \si(\rho)
\end{equation}
\item The above also provides an explanation for the counter
example to MinEP given by Landauer, \cite{L1,L2}. There one
considers an electrical circuit with resistance $R$, inductance
$L$, and voltage source $\caE$ in series. The physical entropy
production  is  $\hat\si(j) = \be R j^2$, corresponding to the
Joule heat caused by the current  $j$ through the  resistance $R$.
Apparently, the stationary current $j^* = \caE / R$ does not
coincide with the minimum of the entropy production.

To understand the situation, we embed the network dynamics in a
stochastic process by combining Kirchhoff's second law with the
Johnson-Nyquist noise voltage on the resistance to get the
equation
\begin{equation}\label{rllange}
  \id j_t = \frac{1}{L}\bigl(\caE - R j_t\bigr)\,\id t +
   \sqrt\frac{2 R}{\be L^2}\,\id W_t
\end{equation}
(the Nyquist prefactor for the noise being determined from the
fluctuation-dissipation relation). That is a linear Langevin
equation of the form~\eqref{rllang} for the current which is odd
under time-reversal. Hence, the conclusion of the previous remark
applies and, in particular, both Theorem~\ref{thm: main} and MinEP
are no longer valid.  Yet, we can obtain the correct variational
principle from the DV-functional.\\
Consider indeed the functional
\begin{equation}
  \bar I(\bar j) = \inf_\mu \{ I(\mu) \rel \langle j \rangle_\mu  = \bar j\}
\end{equation}
which, by the contraction principle, is the large deviation rate
function for the empirical average $j_T = \frac{1}{T}\int_0^T
j_t\,\id t$ as $T \uparrow +\infty$,
\begin{equation}\label{rc}
 \bsP[j_T \simeq \bar j] \propto \exp [-T \bar I(\bar j)]
\end{equation}
Here it is easy to compute from \eqref{eq: diffusion-I}:
\begin{equation}
  \bar I(\bar j) = \frac{\be R}{4} \Bigl( \bar j - \frac{\caE}{R} \Bigr)^2
\end{equation}
and that is then also the corrected variational functional to
consider.\\
 For other examples and for further details we refer to
\cite{BMN}.
%where
%$R$ is the resistance in series with inductance
%$L$. The equation
%\eqref{rllang} expresses Kirchhoff's second law for the current $I$.  The noise is physically
%arising from the Johnson-Nyquist effect, with prefactor determined from the
%fluctuation-dissipation relation. We thus have again a linear Langevin equation, but the
%fluctuations concern the current $I_t$ which is odd under time-reversal.
% The correct expression for the
%expected entropy production is here
%\[
%\si(I) = \beta RI^2
%\]
%Hence, the most probable current $I^* = \caE/R$ does not correspond to a minimization of the
%entropy production. The correct variational functional $\caJ(I)$  for the stationary current is
%rather
%\begin{equation}\label{jrl}
%  \caJ(I) = \frac{\be R}{4}
%  \bigl(I - \frac{\caE}{R}\bigr)^2
%  \end{equation}
%and is obtained by computing the large deviations of $i_T =
%\frac{1}{T} \int_0^T I_t\,\id t,\;T\uparrow +\infty$:
%\begin{equation}\label{rc}
% \bsP[i_T \simeq I] \propto \exp -T \caJ(I)
% \end{equation}
%in the usual sense of level-1 large deviations for the occupation times.  The rate $\caJ$ is the
%direct analogue of the DV-functional \eqref{eq: imu}.  For other examples and for further details
%we refer to \cite{BMN}.
\item Our result as formulated in Theorem~\ref{thm: main} is no
longer valid if we are away from the perturbation regime and the
assumptions are not verified. As an example, consider again a
Markov dynamics on a finite state space $\Om$ and let $\mu$ be a
distribution supported in some $\Om_0 \varsubsetneq \Om$, i.e.,
$\mu(x) = 0$ for all $x \in \Om \setminus \Om_0$. As one
immediately checks from~\eqref{eq: sigma-chain}, $\si(\mu) =
+\infty$ whenever there are some $x \in \Om_0$, $y \in \Om
\setminus \Om_0$ such that $\mu(x) k(x,y) \neq 0$. On the other
hand, the DV-functional is bounded: $I(\mu) \leq \max_x
\sum_{y \neq x} k(x,y)$.
\item The Donsker-Varadhan theory is not
restricted to the time-averages in the sense of \eqref{timea}.
More generally, one can study fluctuations along a discrete
sequence of observations with a time interval $\tau$ between the
observations. The time-averages are then of the form
\[
\frac{\tau}{T}\big(F(X_\tau) + F(X_{2\tau}) +\ldots +
F(X_{T-\tau}) + F(X_{T})\big)
\]
and we are concerned with their large deviations along the limit $T=n\tau \uparrow +\infty$. For
every
$\tau$ there is a rate function
$I_\tau(\mu)$. The case
\eqref{eq: imu} corresponds to
$\lim_{\tau\downarrow 0} I_\tau(\mu)/\tau = I(\mu)$.  Obviously,
one can investigate the
 close-to-equilibrium behavior for
  every one of these cases and,
   in principle, one obtains for each of them
  a variational principle.

  %paragraph deleted.

%is directly and solely expressed in terms of the
%dynamical rules; an opposite extreme is the case
%$\tau = T
%\uparrow +\infty$ where the rate function $I_{+\infty}(\mu) = S(\mu \rel \rho)$ is just the relative
%entropy with the stationary measure --- not too surprising but also not very useful as a
%nonequilibrium characterization.
\end{enumerate}

\vspace{5mm}
\noindent{\bf Acknowledgment}\\
K.~N.~is grateful to the Instituut voor Theoretische Fysica,
K.~U.~Leuven for kind hospitality, and acknowledges the support
from the project AVOZ10100520 in the Academy of Sciences of the
Czech Republic.

% ---------------------------------------------------------------------------------------

% --- References ---
\bibliographystyle{plain}

\end{document}